\documentstyle[12pt]{article}
\begin{document}

\begin{center}
{\bf HEAVY QUARK HADROPRODUCTION IN $k_T$-FACTORIZATION APPROACH
WITH UNINTEGRATED GLUON DISTRIBUTIONS}

\vspace{0.5cm}

 Yu.M.Shabelski and A.G.Shuvaev \\
Petersburg Nuclear Physics Institute, \\
Gatchina, St.Petersburg 188300 Russia \\

\end{center}

\vspace{0.5cm}

\begin{abstract}
We consider the processes of compare the heavy quark production
using the unintegrated gluon distributions. The numerical
predictions for high energy nucle\-on-nucleon and photon-nucleon
collisions of the $k_T$-factorization approach (semihard theory)
are compared with the experimental data from Tevatron-collider and
HERA. The total production cross sections and $p_T$ distributions
are considered and they are in reasonable agreement with the data
for rather value of QCD scale.

\end{abstract}

\vspace{2cm}

E-mail SHABELSK@THD.PNPI.SPB.RU

E-mail SHUVAEV@THD.PNPI.SPB.RU

\newpage

\section{Introduction}

The investigation of heavy quark production in high energy hadron
collisions is an important method for studying the quark-gluon
structure of hadrons. The description of hard interactions in hadron
collisions within the framework of QCD is possible only with the help of
some phenomenology, which reduces the hadron-hadron interaction to the
parton-parton one via the formalism of the hadron structure functions.
The cross sections of hard processes in hadron-hadron interactions can
be written as the convolutions squared matrix elements of the
sub-process calculated within the framework of QCD, with the parton
distributions in the colliding hadrons.

The most popular and technically simplest approach is the so-called QCD
collinear approximation, or parton model (PM). In this model all
particles involved are assumed to be on mass shell, carrying only
longitudinal momenta, and the cross section is averaged over two
transverse polarizations of the incident gluons. The virtualities $q^2$
of the initial partons are taken into account only through their
structure functions. The cross sections of QCD subprocess are calculated
usually in the next to leading order (NLO) of $\alpha_s$ series
\cite{1,NDE,Beer,Beer1}. The transverse momenta of the incident partons
are neglected in the QCD matrix elements. This is the direct analogy of
the Weizsaecker-Williams approximation in QED.

Another possibility to incorporate the incident parton transverse
momenta is referred to as $k_T$-factorization approach
\cite{CCH,CE,MW,CH,CC,HKSST}, or the theory of semihard interactions
\cite{GLR,LR,8,lrs,3,SZ,SS,BS,BZ,LSZ,BLZ}. Here the Feynman diagrams
are calculated taking account of the virtualities and of all possible
polarizations of the incident partons. In the small $x$ domain there
are no grounds to neglect the transverse momenta of the gluons,
$q_{1T}$ and $q_{2T}$, in comparison with the quark mass and
transverse momenta, $p_{iT}$.  Moreover, at very high energies and
very high $p_{iT}$ the main contribution to the cross sections comes
from the region of $q_{1T} \sim p_{1T}$ or $q_{2T} \sim p_{1T}$, see
\cite{our,our1,our2} for details. The QCD matrix elements of the
sub-processes are rather complicated in such an approach. We have
calculated them in the LO. On the other hand, the multiple emission of
soft gluons is included here. That is why the question arises as to
which approach is more constructive.

In Sect.~2 we present the cross sections of heavy quark production in
hadron-nucleon and photon-nucleon collisions in the
$k_T$-factorization approach. This approach accounts for the diagrams
without neglecting by the virtualities of incident partons (mainly
gluons), their polarization, etc. The detailed discussion of the
$k_T$-factorization approach, and its numerical difference from the
parton model can be found in \cite{our1,our2}.

The different approaches for the unintegrated gluon distributions are
discussed in Sect.~3 and 4. We show that the difference between them is
not very large, and after integration the unintegrated gluon
distributions reproduce the initial structure functions. At the moment
we have no more realistic unintegrated parton distributions which are
extracted from the data by the same way as structure functions, i.e. by
using the evolution equations, boundary conditions at some small $Q^2_0$,
etc.

The experimental data on heavy quark production at Tevatron-collider
and HERA are discussed in Sect.~5. Our predictions with the selected
$k_T$-factorization (which seens to be rather natural) are in reasonable
agreement with $p_T$-distributions of $b$-quarks produced in FNAL, as
well as with the data on total cross sections of charm and beauty
production at HERA.  Predictions of the $k_T$-factorization approach for
heavy quark production at LHC are also given.

\section{Heavy quark production in hadron-hadron and photon-hadron
collisions in $k_T$-factorization approach}

The conventional parton model expression for the calculation of
heavy quark hadroproduction cross sections has the factorized form
\cite{CSS}:
\begin{equation}
\sigma (a b \rightarrow Q\overline{Q}) = \sum_{ij} \int dx_i dx_j
g_{a/i}(x_i,\mu_F) g_{b/j}(x_j,\mu_F) \hat{\sigma} (i j
\rightarrow Q \overline{Q}) \;, \label{pm}
\end{equation}
where $g_{a/i}(x_i,\mu_F)$ and $g_{b/j}(x_j,\mu_F)$ are the structure
functions of partons $i$ and $j$ in the colliding hadrons $a$ and $b$,
$\mu_F$ is the factorization scale (i.e. virtualities of incident
partons) and $d\,\hat{\sigma} (i j \rightarrow Q \overline{Q})$ is the
cross section of the direct subprocess $(i j \rightarrow \bar{Q}Q$
which is calculated in perturbative QCD. The last cross section can be
written as a sum of LO and NLO contributions, and its analitical form
can be found in \cite{1,NDE,Beer,Beer1}.

The principal problem of parton model is the collinear approximation.
The transverse momenta of the incident partons, $q_{iT}$ and $q_{jT}$
are assumed to be zero, and their virtualities are accounted through the
structure functions only; the cross section
$\hat{\sigma} (i j \rightarrow Q \overline{Q})$ is assumed to be
independent on these virtualities. Naturally, this approximation
essentially simplifies calculations.

Sometimes namely these simplifications result to the serious
disagreement with the experimental data.

Without the discussed simplifications of PM, the differential cross
section of heavy quark hadroproduction in LO QCD (Fig.~1) has the
following form \cite{GLR,lrs}:\footnote{We put the argument of
$\alpha_s$ to be equal to gluon virtuality, which is very close to the
BLM  scheme\cite{blm}; (see also \cite{lrs}).}
\begin{eqnarray}
\frac{d\sigma_{pp}}{dy^*_1 dy^*_2 d^2 p_{1T}d^2
p_{2T}}\,&=&\,\frac{1}{\pi^2} \frac{1}{(s^\prime)^2}\int\,d^2 q_{1T}
d^2 q_{2T} \delta (q_{1T} + q_{2T} - p_{1T} - p_{2T}) \nonumber \\
\label{spp} &\times &\,\frac{\alpha_s(q^2_1)}{q_1^4}
\frac{\alpha_s (q^2_2)}{q^4_2} f_g(y,q_{1T},\mu)
f_g(x,q_{2T},\mu) \vert M_{QQ}\vert^2.
\end{eqnarray}

Here we have used the Sudakov (light cone)
decomposition for quark momenta $p_{1,2}$ through the momenta of
colliding hadrons  $p_A$ and $p_B$,
\begin{equation}
\label{pApB} p^2_A = p^2_B \simeq 0, \qquad 2\,p_A\cdot
p_B\,=\,s^\prime,
\end{equation}
and transverse ones $p_{1,2T}$:
\begin{equation}
\label{p12} p_{1,2} = x_{1,2} p_B + y_{1,2} p_A + p_{1,2T}.
\end{equation}

In LLA kinematics
\begin{eqnarray}
q_1\, \simeq \,yp_A + q_{1T}, && q_2 \simeq \,xp_B +
q_{2T},\nonumber \\ \label{q1q2} q_1^2\, \simeq \,- q_{1T}^2, &&
q_2^2 \simeq \,- q_{2T}^2,  \\ x\,=\,x_1\,+\,x_2, &&
y\,=\,y_1\,+\,y_2. \nonumber
\end{eqnarray}
(The more accurate relations are $q_1^2 =- q_{1T}^2/(1-y)$, $q_2^2
=- q_{2T}^2/(1-x)$ but we are working in the kinematics where $x,y
<< 1$).

% $s^\prime = 2p_A p_B\,\,$,
$q_{1,2T}$ are the gluon transverse momenta;

\begin{eqnarray}
x_1 \,=\, \frac{m_{1T}}{\sqrt{s^\prime}}\, e^{-y^*_1} &,& y_1 \,=\,
\frac{m_{1T}}{\sqrt{s^\prime}}\, e^{y^*_1}, \nonumber \\
\label{xym}
x_2\,=\,\frac{m_{2T}}{\sqrt{s^\prime}}\, e^{-y^*_2} &,&
y_2\,=\,\frac{m_{2T}}{\sqrt{s^\prime}}\, e^{y^*_2},   \\
m_{1,2T}^2\, &=& \,m^2\,+\,p_{1,2T}^2,\nonumber
\end{eqnarray}
where $y^*_{1,2}$  are the quark rapidities in the hadron-hadron
c.m.s. frame and $m$ is the mass of produced heavy quark and we
take these masses
\begin{equation}
m_c = 1.4\; GeV\;\;, m_b = 4.6\; GeV \;,
\end{equation}
for the values of short-distance perturbative quark masses
\cite{Nar,BBB}. $\vert M_{QQ}\vert^2$ is the square of the matrix
element for the heavy quark pair hadroproduction. It is calculated in
the Born approximation of QCD without standard simplifications of the
parton model. Contrary to the mention of \cite{BS}, the
transformation Jacobian from $x, y$ to $y^*_1, y^*_2$ is accounted
in our matrix element.

The unintegrated gluon distributions $f_g(y,q_{1T},\mu)$ and
$f_g(x,q_{2T},\mu)$ will be discussed in details in the Sect. 3.

To provide the $k_T$ factorization of eq.(2)
 the cross-section of the elementary subprocess for the heavy quarks
production, $gg \to Q \overline{Q}$ was calculated in the axial gauge
 where the spin part
of the gluon ($q_i$) propagator takes  the form
$$ d_{\mu\nu}(q_i)=g_{\mu\nu}-\frac{n_\mu q_{i,\nu}+q_{i,\mu} n_\nu}
{(n\cdot q_i)}$$
 with the gauge vector $n_\nu=(p_1+p_2)_\nu -
(p_1+p_2)_{T,\nu}$ (that is $n_\nu=(n_0,0,0,0)$ in the frame where the
longitudinal part of quark pair momentum $(p_1+p_2)_z=0$).

In this gauge both the DGLAP and BFKL leading logarithms in the
unintegrated distributions $f_g$ are given by the ladder type diagrams
shown in Fig.1 without any extra interactions between the upper and
lower parts of the graphs (that is between the functions
$f_g(x,q_{2T},\mu)$ and $f_g(y,q_{1T},\mu)$).

Being averaged over the
colours of the incomimg gluons
 the matrix element squared $|M_{QQ}|^2$ reads

\newpage

\begin{eqnarray} |\,M_{QQ}|^2\,&=&\,\frac 1{x^2y^2}\left\{\,\frac
1{4N_c}\,
\left[\,\frac{m_{11}}{(m^2-t)^2}\,+\,\frac{m_{22}}{(m^2-u)^2}\,\right]\right.
\nonumber \\ \label{M^2}
&-&\,\frac 1{4N_c}\,\frac{(m_{12}+m_{21})/2}{(m^2-t)(m^2-u)} \\
&+&\,\left.\frac 12\,N_c(N_c^2-1)^{-1}\,\frac 1s
\left[\frac{(m_{13}+m_{31})/2}{m^2-t}\,-\,
\frac{(m_{23}+m_{32})/2}{m^2-u}\,+\,\frac{m_{33}}{s}
\right]\,\right\}. \nonumber
\end{eqnarray}
Here $N_c=3$ is the number of colors, the variables $s$, $t$, $u$
are defined for $gg \to Q \overline{Q}$ subprocess.

\begin{eqnarray}
m^2-t\,&=&\,m^2\,+\,x_1y_2s^\prime\,+\,(p_1-q_1)_T^2 \nonumber \\
m^2-u\,&=&\,m^2\,+\,x_2y_1s^\prime\,+\,(p_1-q_2)_T^2 \nonumber \\
s\,&=&\,xys^\prime\,-\,(p_1+p_2)_T^2 \nonumber
\end{eqnarray}
The elements $m_{ik}$ are the numerators of the lowest order QCD Born
diagrams squared and averaged over quark helicities; the incoming
transverse momenta $q_{1,2T}$ play the role of gluon
polarization vectors,
\begin{eqnarray}
m_{11}\,&=&\,{\rm Sp}\,(\hat p_1+m)\,\hat q_{1T}\,(\hat p_1-\hat
q_1+m)\, \hat q_{2T}\,(\hat p_2-m)\,\hat q_{2T}\,(\hat p_1-\hat
q_1+m)\,\hat q_{1T} \nonumber \\ m_{21}\,&=&\,{\rm Sp}\,(\hat
p_1+m)\,\hat q_{1T}\,(\hat p_1-\hat q_1+m)\, \hat q_{2T}\,(\hat
p_2-m)\,\hat q_{1T}\,(\hat p_1-\hat q_2+m)\,\hat q_{2T} \nonumber
\\ m_{22}\,&=&\,{\rm Sp}\,(\hat p_1+m)\,\hat q_{2T}\,(\hat
p_1-\hat q_2+m)\, \hat q_{1T}\,(\hat p_2-m)\,\hat q_{1T}\,(\hat
p_1-\hat q_2+m)\,\hat q_{2T} \nonumber \\ m_{31}\,&=&\,{\rm
Sp}\,(\hat p_1+m)\,\hat V_3\, (\hat p_2-m)\,\hat q_{2T}\,(\hat
p_1-\hat q_1+m)\,\hat q_{1T} \nonumber \\ m_{32}\,&=&\,{\rm
Sp}\,(\hat p_1+m)\,\hat V_3\, (\hat p_2-m)\,\hat q_{2T}\,(\hat
p_1-\hat q_2+m)\,\hat q_{2T} \nonumber \\ m_{33}\,&=&\,{\rm
Sp}\,(\hat p_1+m)\,\hat V_3\,(\hat p_2-m)\,\hat V_3, \nonumber
\end{eqnarray}
$$ m_{12}\,=\,m_{21},\;\;m_{13}\,=\,m_{31},\;\; m_{23}\,=\,m_{32}.
$$ The matrix $$ \hat V_3 \,=\, q_{1T}\cdot q_{2T}\,(\hat q_1-\hat
q_2) \,- \,\bigl((2q_1+q_2)\cdot q_{2T}\bigr)\,\hat q_{1T}
\,+\,\bigl((2q_2+q_1)\cdot q_{1T}\bigr)\,\hat q_{2T} $$ is the
contribution of a triple gluon vertex.

Thus we can calculate the heavy quark production cross section without
standart simplifications of the parton model since in the small $x$
region there are no grounds for neglecting the transverse momenta of
the gluons $q_{1,2T}$ in comparison with the quark mass and transverse
momenta $p_{1,2T}$. Although explicit expressions for $m_{ik}$ are
rather bulky, they can be easily computed using any analytical computer
program (say, REDUCE or MATHEMATICA). For example, the dependence of the
matrix elements on the gluon transverse momenta is a main source for the
nontrivial azimuthal correlations arising between quarks momenta
$p_{1,2T}$ which could not be described in the conventional parton
model.

In the case of high energy heavy quark photoproduction on the proton
target we should consider two possibilities, the same resolved
production via quark-gluon structure of the photon, Fig.~1, where one
proton should by replaced by photon (with its parton distribution,
different from the case of incident nucleon), and direct production in
the photon-gluon fusion, $\gamma p \to Q\bar{Q}$, Fig.~2,  where the
incident photon has the fixed momentum.

The cross section of the direct interaction can be written by the same
way as Eq. (2) where the unintegrated gluon distribution
$f_g(y,q_{1T},\mu)$ should be replaced by $\delta(y - 1)$.
The cross section of the direct interaction can be written as
\begin{eqnarray}
\frac{d\sigma_{\gamma p}}{d^2p_{1T}} & = &
\frac{\alpha_{em}e^2_Q}{\pi} \int dz d^2q_T
\frac{f_g(x,q_T,\mu)}{q^4_T} \alpha_s(q^2) \nonumber \\ & \times &
\left\{ [(1-z)^2+z^2] \left(\frac{\vec{p}_{1T}}{D_1} +
\frac{\vec{q}_T-\vec{p}_{1t}}{D_2} \right )^2 + m^2_Q
\left(\frac1{D_1} - \frac1{D_2} \right )^2 \right\} \;,
\end{eqnarray}
\begin{equation}
D_1 = p^2_{1T} + m^2_Q \;,\;\; D_2 = (\vec{q}_T - \vec{p}_{1T})^2
+ m^2_Q \;,
\end{equation}
here $\alpha_{em} = 1/137$ and $e_Q$ is the electric charge of the
produced heavy quark.

\section{Unintegrated gluon distributions in different approaches}

 The unintegrated parton distribution
$f_a(x,q_T,\mu)$ determines the probability to find a parton $a$
initiating the hard process with the transverse momentum $q_T$
(and with factorization scale $\mu$).

At very low $x$, that is to leading $\log(1/x)$ accuracy,
for the case of $q_T$ of the order of scale $\mu$ the
unintegrating gluon distribution are approximately determined
\cite{GLR} via the derivative of the usual structure function:
\begin{equation}
\label{xG}
 f_g(x,q_T,\mu)\ \simeq f_g(x,q^2)\ =\
\frac{\partial[xg(x,q^2)]} {\partial \ln q^2}\ .
\end{equation}
However at a large $x$ the unintegrated density (\ref{xG}) becomes
 negative.\\

To restore the function
$f_a(x,q_T,\mu)$ on the basis of the conventional (integrated)
parton density $a(x,\lambda^2)$ we have to consider the DGLAP
evolution\footnote{For the $g \to gg$ splitting we need to insert
a factor $z'$ in the last integral of Eq. (12) to account for the
identity of the produced gluons.}
\begin{equation}
\frac{\partial a}{\partial \ln \lambda^2} = \frac{\alpha_s}{2 \pi}
\left[ \int^{\Delta}_x P_{aa'} (z) a'(\frac xz,\lambda^2) dz -
a(x,\lambda^2) \sum_{a'} \int^{\Delta}_0 P_{a'a} (z') dz'
\right] \;.
\end{equation}
Here, $a(x,q^2_1)$ denotes $xg(x,q^2_1)$ or $xq(x,q^2_1)$, and
$P_{aa'}$ are the splitting functions.

The first term on the right-hand side of Eq. (12) describes the
number of partons $\delta_a$ emitted in the interval $\lambda^2 <
q^2_T < \lambda^2+\delta \lambda^2$, while the second (virtual)
term reflects the fact that the parton $a$ disappears after the
splitting.

The second contribution may be resummed to give the survival
probability $T_g$ that gluon $g$ with the transverse momentum
$q_T$ remains untouched in the evolution up to the factorization
scale
\begin{equation}
\label{Sud} T_g(q_T,\mu) = \exp \left[ -\int^{\mu^2}_{q^2_T}
\frac{\alpha_s(p_T)}{2\pi} \frac{dp^2_T}{p^2_T} \sum_{a'}
\int^{\Delta}_0 P_{a'g} (z') dz' \right] \;.
\end{equation}

Thus, the unintegrated gluon distribution $f_g(x,q_T,\mu)$ has the
form
\begin{equation}
f_g(x,q_T,\mu) = \sum_{a'}\left[\frac{\alpha_s}{2 \pi}
\int^{\Delta}_x P_{ga'} (z) a'\left(\frac xz,q_T^2 \right) dz
\right] T_g(q_T,\mu) \;,
\end{equation}
where the cut-off $\Delta$ is used in Eqs.~(12-14) \cite{DDT,KMR}.

The expression (14) with the survival probability (13) provides the
positivity of the unintegrated probability $f_g(x,q_T,\mu)$ in the
whole interval $0 < x < 1$. It is necessary to note that the
calculation of the integrals in Eqs.~(13) and (14) should be fulfilled
with rather high and equal accuracy because the factor $T_g(q_T,\mu)$
compensate the singularity at $\Delta \to 1$ in Eq.~(14).

For the case of one loop QCD running coupling $\alpha_s(p_T) =
4\pi /(b \ln{p^2_T/\Lambda^2})$ the factor $T_a(q_T,\mu)$ can be
written down explicitely. In particular, the gluon survival
probability (which enters our formulae) reads:
\begin{eqnarray}
& T_a ( q_T, \mu ) = \frac{\ln(\mu /\Lambda)}{\ln(q_T /\Lambda)}
\cdot \exp \left\{ \frac{8N_c}{b} \left[ \ln \frac{\mu}{q_T} - \ln
\frac{\mu}{\Lambda} \ln \frac{\ln(\mu /\Lambda
)}{\ln(q_T/\Lambda)} - 2E_1(q_T,\mu) + 3/2 E_2(q_T,\mu) - \right.
\right. \nonumber \\ \nonumber \\ & \left. -  2/3 E_3(q_T,\mu) +
1/4 E_4(q_T,\mu) \right] + \\ \nonumber \\ &\left. +
\frac{2}{3b}n_F \left[ 3E_1(q_T,\mu) - 3E_2(q_T,\mu) +
2E_3(q_T,\mu) \right] \right\} \nonumber
\end{eqnarray}
where
\begin{equation}
E_k(q_T,\mu) = \left( \frac{\Lambda}{\mu} \right)^k
[Ei(k\ln(\mu/\Lambda )) - Ei(k\ln(q_T/\Lambda )) ] \;,
\end{equation}
and the integral exponent $Ei(z) =
\int^z_{-\infty}\frac{dt}{t}\exp{t}$; $n_F$ and $N_c$ are the
number of light quark flavoures and the number of colours,
respectively, and $b = 11 - \frac23 n_F$.

In the leading $\log (1/x)$ (i.e. BFKL) limit the virtual DGLAP
contribution is neglected. The scale $\mu$ is of the order of $q_T$.
 So $T_a = 1$ and one comes back to
Eq.~(11)
\begin{equation}
f_a^{BFKL} (x,q_T,\mu) = \frac{\partial a(x,\lambda^2)} {\partial
\ln \lambda^2}\;\;\;, \lambda = q_T \; .
\end{equation}

In the double log limit Eq. (12) can be written in the form
\begin{equation}
f_a^{DDT}(x,q_T,\mu) = \frac\partial{\partial \ln \lambda^2}
\left[a(x,\lambda^2) T_a(\lambda,\mu)\right]_{\lambda = q_T} \; ,
\end{equation}
which was firstly proposed by \cite{DDT}. In this limit the
derivative $\partial T_a /\partial \ln \lambda^2$ cancels the
second term of the r.h.s. of Eq. (12) (see \cite{KMR} for a more
detailed discussion)\footnote{There is a cancellation between the
real and virtual soft gluon DL contributions in the DGLAP
equation, written for the integrated partons (including all
$k_T\le\mu$). The emission of a soft gluon with momentum fraction
$(1-z)\to 0$ does not affect the $x$-distribution of parent
partons. Thus the virtual and real contributions originated from
$1/(1-z)$ singularity of the splitting function $P(z)$ cancel each
other. On the contrary, in the unintegrated case the emission of
soft gluon (with $q'_T>k_T)$ alters the transverse momentum of
parent ($t$-channel) parton.}.

Finally, the probability $f_a(x,q_T,\mu)$ is related to the BFKL
function $\varphi(x,q^2_T)$ by
\begin{equation}
\varphi (x,q^2_T)\ =\ 4\sqrt2\,\pi^3 f_a(x,q_T,\mu)/q^2_T \; .
\end{equation}

Note that due to a virtual DGLAP contribution the derivative
$\partial a(x,\lambda^2) / \partial \lambda^2$ can be negative for
not small enough $x$ values. This shortcoming of Eq.~(17) is
partly overcome in the case of Eq.~(18).

We have to emphasize that $f_g(x,q_T,\mu)$ is just the quantity
which enters into the Feynman diagrams. The distributions
$f_g(x,q_T,\mu)$ involve two hard scales\footnote{This property is
hidden in the conventional parton distributions as $q_T$ is
integrated up to the scale $\mu$.}: $q_T$ and the scale $\mu$ of
the probe. The scale $\mu$ plays a dual role. On the one hand it
acts as the factorization scale, while on the other hand it
controls the angular ordering of the partons emitted in the
evolution \cite{MCi,CFM,Mar}. The factorization scale $\mu$
separates the partons associated with the emission off both the
beam and target protons (in $pp$ collisions) and off the hard
subprocess. For example, it separates emissions off
%***** not sure "off' or 'of'
the beam (with polar angle $\theta < 90^o$ in c.s.m.) from those
off the target (with $\theta > 90^o$ in c.s.m.), and from the
intermediate partons from the hard subprocess. This separation was
proved in \cite{MCi,CFM,Mar} and originates from the destructive
interference of the different emission amplitudes (Feynman
diagrams) in the angular boundary regions.

If the longitudinal momentum fraction is fixed by the hard
subprocess, then the limits of the angles can be expresseed in
terms of the factorization scale $\mu$ which corresponds to the
upper limit of the allowed value of the $s$-channel parton $k_T$.

\section{Numerical values of unintegrated gluon distributions}

As was shown above, there exist several ways for estimation the gluon
unintegrated distributions. Now we will compare numerically several of
them to see the theoretical uncertainties which should result in the
uncertainties in the predictions for cross sections of heavy quark
production.

First of all, it is evident that unintegrated gluon distributions
obtained with the help of Eqs.~(12)-(14) should coincide after their
integration over $q^2$ with the used structure functions,
\begin{equation}
xg(x,q^2) = \int \frac{dq^2}{q^2} f_g(x,q_T,\mu) \;.
\end{equation}

Some problem is that the values of $f_g(x,q_t,\mu)$ are principally
unknown at very small $q_T^2 \simeq q^2$. So instead of Eq.~(20) we
can write
\begin{equation}
xg(x,q^2) = xg(x,q^2_0) + \int_{q^2_0} \frac{dq^2}{q^2} f_g(x,q_T,\mu)
\;.
\end{equation}
with $q^2_0 \sim$ 1 GeV$^2$.

The calculated values of heavy quark production cross sections depend
\cite{ShS} on the analytical form of the cut-off parameter $\Delta$ in
Eqs.~(12)-(14). However, if Eqs.~(13) and (14) are calculated with the
same numerical accuracy, this dependence becomes rather weak.

Now we compare the results for sum rules, Eq.~(21), for two cases,
\begin{equation}
\Delta = \frac{\mu}{\mu + q_T}
\end{equation}
and
\begin{equation}
\Delta = \frac{\mu}{\sqrt{\mu^2 + q_T^2}} \;.
\end{equation}
The results of calculations are presented in Fig. 3a and one can see
that the variants (22) and (23) give practically the same results. In
the futher numerical calculations we will use the variant (22) which
corresponds to the angular ordering in the gluon emission
\cite{MCi,CFM,Kw}.

In Fig.~3b we compare the $q^2$ depencences of unintegrated gluon
distributions given by Eq.(14) \cite{KMR}, the KMS distribution
\cite{KMS}, which was based on the BFKL equation and fitted to the
$F_2$ HERA data just in unintegrated form, and simplified formula (11)
\cite{GLR}. In the first case we present the results for the value of
QCD scale $\mu^2$ in Eqs. (13), (14) equal to 100 GeV$^2$ (i.e. about
$4m_b^2$). Here we have used rather more or less realistic gluon
distribution GRV94 HO \cite{GRV} which is compatible with the most
recent data, see discussion in Ref.~\cite{GRV1}. For a low $x$ all
three distributions are rather close to each other while at a larger
$x \sim 0.05$ the KMR prescription gives for $q^2>10$ GeV$^2$ about
twice larger gluon density.

Eq.~(2) enables us to calculate straightforwardly all
distributions concerning one-particle or pair production.
One-particle calculations as well as correlations between two
produced heavy quarks can be easily done using, say, the VEGAS
code \cite{Lep}.

However there exists a principle problem coming from the infrared
region.  Since the functions $\varphi (x,q^2_i)$ as well as
$f_g(x,q^2_{Ti},\mu)$ are unknown at small values of $q^2_2$ and
$q^2_1$, i.e. in nonperturbative domain we calculate separately
the contributions from $q^2_1 < Q^2_0$, $q^2_2 < Q^2_0$, $q^2_1 >
Q^2_0$ and $q^2_1 > Q^2_0$ \cite{our,our1,our2}.

\newpage

$$\int d^2 q_{1T} d^2 q_{2T} \delta (q_{1T} + q_{2T} - p_{1T} -
p_{2T}) \frac{\alpha_s(q^2_1)}{q_1^4} \frac{\alpha_s
(q^2_2)}{q^4_2} f_g(y,q_{1T},\mu) f_g (x,q_{2T},\mu) \vert
M_{QQ}\vert^2 = $$
\begin{equation}
\label{int} =  \alpha^2_s (Q^2_0)  \,
xg(x,Q^2_0)\, yg(y,Q^2_0)\, T^2(Q_0^2,\mu^2)\, \left (\frac{\vert
M_{QQ} \vert^2}{q^2_1 q^2_2} \right)_{q_{1,2}\rightarrow 0} \; +
\end{equation}
$$ + \;  \alpha_s (Q^2_0)
xg(x,Q^2_0)\,T(Q_0^2,\mu^2)\, \int^{\infty}_{Q^2_0} \, d
q^2_{1T}\, \delta (q_{1T} - p_{1T} - p_{2T})\,\, \times $$ $$
\times \, \frac{\alpha_s (q^2_1)}{q^4_1} f_g(y,q_{1T},\mu) \left
(\frac{\vert M_{QQ} \vert^2}{q^2_2} \right)_{q_2\rightarrow 0} \;
+ $$ $$ + \;  \alpha_s (Q^2_0)\,
yg(y,Q^2_0)\,T(Q_0^2,\mu^2)\, \int^{\infty}_{Q^2_0} \, d
q^2_{2T}\, \delta (q_{2T} - p_{1T} - p_{2T})\, \, \times $$ $$
\times \, \frac{\alpha_s (q^2_2)}{q^4_2} f_g(x,q_{2T},\mu) \left
(\frac{\vert M_{QQ} \vert^2}{q^2_1} \right)_{q_1 \rightarrow 0} \;
+ $$ $$ + \; \int^{\infty}_{Q^2_0} \, d^2 q_{1T}
\int^{\infty}_{Q^2_0} \, d^2 q_{2T}\, \delta (q_{1T} + q_{2T} -
p_{1T} - p_{2T}) \, \times $$ $$ \times
\,\frac{\alpha_s(q^2_1)}{q_1^4} \frac{\alpha_s (q^2_2)}{q^4_2}
f_g(y,q_{1T},\mu) f_g (x,q_{2T},\mu) \vert M_{QQ}\vert^2 \;, $$
where the unintegrated gluon distributions $f_g(x,q_T,\mu)$ are
taken from Eq.~(14). In the numerical calculations we use the values
$\mu^2 = m^2_T$, $\mu^2 = 4 m^2_T$ and $\mu^2 = \hat{s}$.

The first contribution in Eq.~(24) ($q^2_1 < Q^2_0$, $q^2_2 < Q^2_0$)
with the matrix element averaged over directions of the two-dimensional
vectors $q_{1T}$ and $q_{2T}$ is exactly the same as the conventional
LO parton model expression. It is multiplied by the 'survival'
probability $T^2(Q_0^2,\mu^2)$ not to have transverse momenta
$q_{1T}, q_{2T} > Q_0$. We assume $Q_0^2$ = 1~GeV$^2$. The sum of the
produced heavy quark momenta is taken to be  exactly zero here.

The next three terms (when one, or both $q^2_{Ti} > Q_0^2$) contain
the corrections to the parton model due to account the different
gluon polarizations, their virtualities and transverse momenta in
the matrix element. The relative contribution of these corrections
strongly depends on the initial energy. If it is not high enough,
the first term dominates, and all results are similar to the
conventional LO parton model predictions \cite{our1}. In the case
of very high energy the opposite situation takes place, the first
term can be considered as a small correction and our results
differ from the conventional ones. So we need the highest energies
for investigation the $k_T$-factorization effects.

%As it was discussed in Sect.~4, the factor $T$ is known with DLog
%accuracy only. When Eq.~(8) gives $f_a < 0$ we put $f_a = 0$. It
%needs small correction only at high energies, where the small-$x$
%region with positive $f_a$ dominates.

\section{Heavy quark production in the $k_T$-factorization approach}

Let us compare the results of our numerical calculations with the
data on beauty production at Tevatron-collider. In Fig. 4a we present
the data \cite{Abb} on $b$ quark production with $p_T > p_{min}$
identified by its muon decay, as the function of $p_{min}$ at
1.8 TeV. The curves of different type show the calculated results
with different scales $\mu^2$ in (13) and (14), $\mu^2 = m_T^2$,
$\mu^2 = 4 m_T^2$ and $\mu^2 = \hat{s}$, respectively,
where $\hat{s}$ is the invariant energy of the produced heavy
quark pair, $\hat{s} = xys$. The calculated values of one-particle
$E_T$ distributions, $d\sigma /dE_T$, in $k_T$-factorization
approach are presented in Fig.~4b together with the data \cite{Abb1}
extracted from muon tagged jets. In both Figs.~4a and 4b our results
are slightly smaller than the data.

The similar data on $b\bar{b}$ production with $p_T > p_{min}$
obtained by UA1 (circles) and CDF Coll. (Triangle point) at
$\sqrt{s} = 630$ GeV taken from \cite{Aco} are presented in Fig.~4c
together with the results of our calculations. Again, our results are
in agreement with UA1 and slightly smaller than the CDF data. Some
disagreement between UA1 and CDF experimental data is evident.

The experimental data for charm production \cite{Aco1} at 1.96 TeV
are compared with our calculations in Fig.~5. Now our curves slightly
overestimate the data.

\vskip 10 pt
\begin{center}
{\bf Table 1} The total cross sections of charm and beauty production
in the $k_T$-factorization approach for $\mu^2 = m^2_Q$.

\vskip 20 pt
\begin{tabular}{|c|r|r|r|r|} \hline

 & \multicolumn{2}{c|}{all rapidities} &
\multicolumn{2}{c|}{$|y_Q| < 1$}  \\ \hline

$\sqrt{s}$ & $c\bar{c}$ & $b\bar{b}$ & $c\bar{c}$ & $b\bar{b}$
\\ \hline

14 TeV   & 18.8 mb & 0.92 mb & 4.24 mb & 0.245 mb \\ \hline

1.96 TeV & 2.95 mb & 89.6 $\mu$b & 843 $\mu$b & 33.3 $\mu$b \\ \hline

1.8 TeV  & 2.7 mb & 84.5 $\mu$b & 780 $\mu$b & 30.2 $\mu$b \\ \hline

\hline
\end{tabular}
\end{center}
\vskip 10 pt

The energy dependences of the total cross sections of $c\bar{c}$
and $b\bar{b}$ pair production are presented in Table~1.

As it was mentioned, at comparatively small energies the first term in
Eq.~(24) dominates and the result of $k_T$-factorization approach
practically coinsides with LO parton model prediction. With increasing
of the energy the difference between the considered approaches increases
and at the LHC energy the $k_T$-factorization predictions are about two
times larger than the LO parton model ones (i.e. of the order of NLO
parton model predictions). Generally, the $k_T$-factorization yields a
more strong energy dependences both for $c\bar{c}$ and $b\bar{b}$
production. It can be explained by additional contributions appearing at
very high energies in the $k_T$-factorization approach, see
\cite{our}. The main part of these contributions corresponds to the
configurations where the transverse momentum of heavy quark is balanced
not by another heavy quark but by the $q_T$ of gluons.

Note that we underestimate the $b$-quark cross section but overestimate
 the charm production. This may be caused either by the behaviour of
the gluon densities used in the calculations  (too low density in the
kinematical region ($x\sim 0.01, \; \mu^2 \sim 50$ GeV$^2$)
corresponding to the beauty production at the Tevatron and too high
density at a lower scale and $x$ relevant for the charm production)
or this may indicate some experimental problem in selecting the
$b$-quark events. Not excluded that sometimes the muon or jet from a
charm production (which has a much larger cross section) was treated
as the lepton/jet coming from the beauty.

It is not quite clear what the renormalization scale $\mu_R$ should be
used as the argument of QCD coupling $\alpha_s$ in Eqs.~(2) and (24)
within the LO approximationof $k_T$-factorization approach. As the main
version in the present paper we choose $\mu_R = Q^2_i$, motivated by
BLM prescription \cite{blm}. On the other hand, based on the explicit
one-loop calculation in Ref.~\cite{DDT}, the argument was given in
favour of the running coupling depending on the largest virtuality in
the ladder cell under consideration\footnote{The preexponent factor
$\ln(\mu/\Lambda)/\ln(q_T/\Lambda)$ in Eq.~(15) effectively replaces the
one-loop running coupling $\alpha_s(\mu)$ by the $\alpha_s(q_T)$, close
to the BLM prescription \cite{blm}.}.

By this reason in Fig.~4a we present the curve corresponding to the
$\mu_R = m_T$ (that is in Eq.~(24) we put the coupling $\alpha_s(m_T)$).
For the case of beauty production this choice reduces the predicteed
cross section of about $2 \div 3$ times for $p_{T min} =$ 5--30 GeV/c,
respectively. However for a lighter charm quark using the
$\alpha_s(m_T)$ we obtain the cross section only $1.2 \div 1.5$ times
smaller than that with $\alpha_s(q^2_i)$.

 Heavy quark production at the Tevatron within the framework of
$k_T$-factorization was studied recently in \cite{LLZ}. The authors
used the unintegrated gluon densities obtained from the Linked Dipole
Chain model. With the "standard" gluons they obtained the results
close to our calculations and underestimated the beauty cross section
extracted from the measurement of the muon and jet distributions.
To describe the data it was needed to take so-called "leading" gluons,
that is the version of the model which neglects the light quark
contribution and the non-singular terms in the splitting functions.\\
 We see no possibility to justify such an approach.

On the other hand the new CDF data where the $B$-meson was identified
via the $B\to J/\Psi\ X$ decay mode give a lower cross section.
The last analysis of this new CDF data gives $29.4^{+6.2}_{-5.4} \mu$b
\cite{CFMNR} in agreement with our result ($25 \mu$b) for beauty
production at 1.96 TeV for $|y_b| < 1$.

The energy dependences of the total cross sections of $c\bar{c}$ and
$b\bar{b}$ photoproduction are presented in Fig.~6. The data on charm
(Fig.~6a) and beauty (Fig.~6b) photoproduction cross sections are taken
from \cite{ZEUS,H1,H1a}. All three values, used for the scale give very
similar results which are again slightly lower than the high energy
data. Here we show separately the sum of direct and resolved
contributions, as well as the only resolved ones. The resolved
contributions have more strong energy dependence. At energies
$W \sim 300$ GeV they give $12-15$\% of the total cross sections but in
some kinematical regions they can even dominate \cite{ShS}, for example,
in the target fragmentation region. The resolved contributions decrease
with $p_T$ more fast, that is connected with the finite phase space
value.

\section{Conclusion}

We have compared the $k_T$-factorization approach for heavy quark
hadro- and photopro\-duction at collider energies with the existing
experimental data. The agreement is reasonable when we use the cut-off
(22) in Eqs.~(13) and (14) and it does not practically depends on the
value of QCD scale $\mu$ in these equations. We present also some
predictions for the total cross sections of heavy quark production at
LHC energies.

Another example of very successful comparison of experimental data
with the $k_T$-factorization approach can be found in \cite{HKSST}
where some different assumptions were used.

It has been shown in Sect.~4 of \cite{our} that the contribution of
the domain with strong $q_T$ ordering ($q_{1T}, q_{2T} \ll
m_T=\sqrt{m_Q^2+p_T^2}$) coincides in the $k_T$-factorization
approach with the LO PM prediction. Besides this a numerically
large contribution appears at high energies in $k_T$-factorization
approach in the region $q_{1,2T}\ge m_T$. This kinematically
relates to the events where the transverse momentum of heavy quark
$Q$ is balanced not by the momentum of antiquark $\overline Q$ but
by the momentum of the nearest gluon.

  An interesting consequence of such a kinematics is the fact
that even at large transverse momentum $p_{1T} >> m$ the ratio
of the single quark inclusive cross sections for charm and beauty
production $R_{c/b}=(d\sigma_c/d^3p)/(d\sigma_b/d^3p)$ does not tend
to unity. At Tevatron-LHC energies this ratio $R_{c/b}\simeq 1.5$
and only selecting the events where both quarks has a large $p_T$
we obtain the ratio close to unit (the difference $R_{c/b}-1$ does
not exceed a few per cents when both $p_{1T},p_{2T} > 40$ GeV).

The configurations with $q_{1,2T}\ge m_T$ are
 associated with the NLO (or even NNLO, if
both $q_{1T}, q_{2T} \ge m_T$) corrections in terms of the PM with
fixed number of flavours, i.e. without the heavy quarks in the
evolution. Indeed, as was mentioned in \cite{1}, up to 80\% of the
whole NLO cross section originates from the events where the heavy
quark transverse momentum is balanced by the nearest gluon jet.
Thus the large "NLO" contribution, especially at large $p_T$, is
explained by the fact that the virtuality of the $t$-channel (or
$u$-channel) quark becomes small in the region $q_T \simeq p_T$,
and the singularity of the quark propagator $1/(\hat{p} - \hat{q}
- m_Q)$ in the "hard" QCD matrix element,
$M(q_{1T},q_{2T},p_{1T},p_{2T})$, reveals itself.

The double logarithmic Sudakov-type form factor $T$ in the
definition of the unintegrated parton density Eq.~(14) comprises an
important part of the virtual loop NLO (with respect to the PM)
corrections. Thus we demonstrate that the $k_T$-factorization
approach collects already at the LO the major part of the
contributions which play the role of the NLO (and even NNLO)
corrections to the conventional PM.  Therefore we hope that a
higher order (in $\alpha_S$) correction to the $k_T$-factorization
could be rather small.

Recall that our results are rather stable with respect of the
factorization scale $\mu$ variations and with respect to the form of
angular cut-off parameter $\Delta$ in (22, (23).

Another advantage of this approach is that a non-zero transverse
momentum of $Q \overline Q$-system ($p_{T pair}=p_{1T}+p_{2T} =
q_{1T}+q_{2T}$) is naturally achieved in the $k_T$-factorization.
We have calculated in \cite{our1,our2} the $p_{T pair}$ distribution
and compared it with the single quark $p_T$ spectrum. At the low
energies the typical values of $p_{T pair}$ are much lesser than the
heavy quark $p_T$ in accordance with collinear approximation. However
for LHC energy both spectra become close to each other indicating
that the transverse momentum of second heavy quark is relatively
small. The typical value of this momentum depends on the parton
structure functions/densities. It increases with the initial energy
and with the transverse momenta of the heavy quarks, $p_T$. Thus
one gets a possibility to describe a non-trivial azimuthal
correlation without introducing a large "phenomenological"
intrinsic transverse momentum of the partons.

The contribution coming from $q_T > \mu$ region could enhance the flux
of colliding gluons and by this way partly explain the
FNAL-Tevatron puzzle -- data on the cross section of $b\bar{b}$
(or high-$p_T$ prompt photon) production are 2-3 times larger than the
conventional NLO PM QCD predictions \cite{KMR,Abb,Apan}
\footnote{Recall, however, that the analysis \cite{CFMNR} of the
new CDF data gives a lower beauty production cross section.}.

At the moment we have no realistic unintegrated parton distributions
which fit the data with accounting for the contribution from
$q_T > \mu$. We hope that future experiments will allow to distinct
between different approaches. It is necessary to note is that the
theoretical results concern the heavy quarks rather than the
hadron production which can be investigated experimentally. The
hadronization leads to several important effects, however their
description needs additional phenomeno\-logical assumptions, see
e.g. \cite{Likh,jdd,Shab}.

We are grateful to M.G.Ryskin for many useful discussions. This paper
was supported in part by grants RCGSS-1124.2003.2 and
NATO PDD (CP) PST.CLG 980287.

\newpage
\begin{center}
{\bf Figure Captions}
\end{center}

Fig. 1. Low order QCD diagrams for heavy quark production in $pp$
($p\overline{p}$) collisions via gluon-gluon fusion.

Fig. 2. Low order QCD diagrams for heavy quark production in direct
$\gamma p$ collisions via photon-gluon fusion.

Fig. 3. The unintegrated gluon distributions integrated over $q^2_T$
until $Q^2$ = 100 GeV$^2$ using Eq.~(21) for $\Delta = \mu/(\mu + q_T)$
(dashed curve) and $\Delta = \mu/\sqrt{\mu^2 + q_T^2}$ (dotted curve)
with $\mu^2 = Q^2$ together with gluon structure function GRV94 HO
(solid curve) which was used for the calculations of these unintegrated
distributions using Eq.~(14) (a). The comparison of KMR, Eq.~(14)
(solid curves), GLR, Eq.~(11) (dotted curves) and KMS \cite{KMS}
(dash-dotted curves) unintegrated gluon distributions at $x = 0.0005$
(upper curves) and $x = 0.05$ (lower curves) (b).

Fig. 4. The cross sections for beauty production for $p_T$ higher
than $p_T^{min}$ (a, c) and $d \sigma/dE_T$ (b) in $p\bar{p}$
collisions at 1.8 TeV (a,b) with $\vert y_1 \vert < 1$, and 0.63
TeV $\vert y_1 \vert < 1.5$, (c) and their description by the
$k_T$-factorization approach with unintegrated gluon distribution
$f_g(x,q_T,\mu)$ given by Eq.~(14). The value of $\Delta$ in
Eq.~(13), (14) was taken equal to $\mu/(q_T + \mu)$ with the scale
values $\mu^2 = m_T^2$ (solid curves, $\mu^2 = 4 m_T^2$ (dashed
curves) and $\mu^2 = \hat{s}$ (dotted curves).

Fig. 5. The cross sections for charm production in $p\bar{p}$ collisions
at 1.96 TeV with $\vert y_1 \vert < 1$ and their description by the
$k_T$-factorization approach with unintegrated gluon distribution
$f_g(x,q_T,\mu)$ given by Eq.~(14). The value of $\Delta$ in Eq.~(13),
(14) was taken equal to $\mu/(q_T + \mu)$ with the scale values
$\mu^2 = m_T^2$ (solid curves, $\mu^2 = 4 m_T^2$ (dashed curves) and
$\mu^2 = \hat{s}$ (dotted curves).

Fig. 6. Total cross sections of charm (a) and beauty (b)
photoproduction. The values of $\Delta$ in Eq.~(13), (14) were taken
equal to $\mu/(q_T + \mu)$ with scale values $\mu^2 = m_T^2$ (solid
curves), $\mu^2 = 4 m_T^2$ (dotted curves) and $\mu^2 = \hat{s}$ (dashed
curves). The resolved photon contributions are shown by lower solid
curves.

\end{document}